\documentclass[conference]{IEEEtran}
\usepackage{multicol}
\usepackage{lipsum,graphicx}
\usepackage{float}
\usepackage{epstopdf}
\usepackage{ifthen}
\usepackage[latin1]{inputenc}

\usepackage{amssymb}
\usepackage[cmex10]{amsmath}
\usepackage{dcolumn}
\usepackage[nolist]{acronym}

\DeclareGraphicsExtensions{.pdf,.jpeg,.png,.eps}
\usepackage{booktabs}

\usepackage[caption=false,font=footnotesize]{subfig}
\usepackage{comment}
\usepackage{color}
\usepackage[table]{xcolor}
\usepackage{cite}

\usepackage{setspace}
\usepackage{mathrsfs}          
\usepackage{amsmath, amssymb, amsthm}
\usepackage{amsfonts}
\usepackage{mathrsfs}
\usepackage{multirow}        
\usepackage{framed}
\usepackage{psfrag}
\usepackage{graphicx}
\usepackage{textcomp, gensymb} 
\usepackage{algpseudocode} 

\newcolumntype{d}[1]{D{.}{\cdot}{#1} }

\usepackage{bm}
\usepackage{mathtools}
\usepackage{siunitx}

\newcommand{\PreserveBackslash}[1]{\let\temp=\\#1\let\\=\temp}
\newcolumntype{C}[1]{>{\PreserveBackslash\centering}p{#1}}
\newcolumntype{R}[1]{>{\PreserveBackslash\raggedleft}p{#1}}
\newcolumntype{L}[1]{>{\PreserveBackslash\raggedright}p{#1}}

\newcommand{\figref}[1]{Fig.~\ref{#1}}

\newcommand{\tabref}[1]{Table~\ref{#1}}

\usepackage{lipsum}
\usepackage{mathtools}
\usepackage{cuted}
\usepackage{stmaryrd}

\usepackage{tabularx}
\usepackage{tabulary}
\usepackage{tikz}
\usepackage{pgfplots}
\usepgfplotslibrary{units}
\pgfplotsset{compat=1.18}

\usepackage{algpseudocode}
\usepackage{algorithm}

\begin{document}
	\begin{acronym}
\acro{1G}{first generation}
\acro{2G}{second generation}
\acro{3G}{third generation}
\acro{3GPP}{Third Generation Partnership Project}
\acro{4G}{fourth generation}
\acro{5G}{fifth generation}
\acro{802.11}{IEEE 802.11 specifications}
\acro{A/D}{analog-to-digital}
\acro{ADC}{analog-to-digital}
\acro{AGC}{automatic gain control}
\acro{AM}{amplitude modulation}
\acro{AoA}{angle of arrival}
\acro{AP}{access point}
\acro{AR}{augmented reality}
\acro{ASIC}{application-specific integrated circuit}
\acro{ASIP}{Application Specific Integrated Processors}
\acro{AWGN}{additive white Gaussian noise}
\acro{BB}{base-band}
\acro{BCJR}{Bahl, Cocke, Jelinek and Raviv}
\acro{BER}{bit error rate}
\acro{BFDM}{bi-orthogonal frequency division multiplexing}
\acro{BPSK}{binary phase shift keying}
\acro{BS}{base stations}
\acro{BCRLB}{bayesian Cram\'{e}r-Rao lower bound}
\acro{CA}{carrier aggregation}
\acro{CAF}{cyclic autocorrelation function}
\acro{Car-2-x}{car-to-car and car-to-infrastructure communication}
\acro{CAZAC}{constant amplitude zero autocorrelation waveform}
\acro{CB-FMT}{cyclic block filtered multitone}
\acro{CCDF}{complementary cumulative density function}
\acro{CDF}{cumulative density function}
\acro{CDMA}{code-division multiple access}
\acro{CFO}{carrier frequency offset}
\acro{CIR}{channel impulse response}
\acro{CM}{complex multiplication}
\acro{COFDM}{coded-\acs{OFDM}}
\acro{CoMP}{coordinated multi point}
\acro{COQAM}{cyclic OQAM}
\acro{CP}{cyclic prefix}
\acro{CRB}{Cramer-Rao bound}
\acro{CPE}{constant phase error}
\acro{CR}{cognitive radio}
\acro{CRC}{cyclic redundancy check}
\acro{CRLB}{Cram\'{e}r-Rao lower bound}
\acro{CS}{cyclic suffix}
\acro{CSI}{channel state information}
\acro{CSMA}{carrier-sense multiple access}
\acro{CWCU}{component-wise conditionally unbiased}
\acro{D/A}{digital-to-analog}
\acro{D2D}{device-to-device}
\acro{DAC}{digital-to-analog converter}
\acro{DBF}{digital beamforming}
\acro{DC}{direct current}
\acro{DFE}{decision feedback equalizer}
\acro{DFT}{discrete Fourier transform}
\acro{DL}{downlink}
\acro{DMT}{discrete multitone}
\acro{DNN}{deep neural network}
\acro{DSA}{dynamic spectrum access}
\acro{DSL}{digital subscriber line}
\acro{DSP}{digital signal processor}
\acro{DTFT}{discrete-time Fourier transform}
\acro{DVB}{digital video broadcasting}
\acro{DVB-T}{terrestrial digital video broadcasting}
\acro{DWMT}{discrete wavelet multi tone}
\acro{DZT}{discrete Zak transform}
\acro{E2E}{end-to-end}
\acro{eNodeB}{evolved node b base station}
\acro{E-SNR}{effective signal-to-noise ratio}
\acro{EVD}{eigenvalue decomposition}
\acro{FBMC}{filter bank multicarrier}
\acro{FD}{frequency-domain}
\acro{FDA}{frequency diverse array}
\acro{FDD}{frequency-division duplexing}
\acro{FDE}{frequency domain equalization}
\acro{FDM}{frequency division multiplex}
\acro{FDMA}{frequency-division multiple access}
\acro{FEC}{forward error correction}
\acro{FER}{frame error rate}
\acro{FFT}{fast Fourier transform}
\acro{FI}{Fisher information}
\acro{FIR}{finite impulse response}
\acro{FM}		{frequency modulation}
\acro{FMT}{filtered multi tone}
\acro{FO}{frequency offset}
\acro{F-OFDM}{filtered-\acs{OFDM}}
\acro{FPGA}{field programmable gate array}
\acro{FSC}{frequency selective channel}
\acro{FS-OQAM-GFDM}{frequency-shift OQAM-GFDM}
\acro{FT}{Fourier transform}
\acro{FTD}{fractional time delay}
\acro{FTN}{faster-than-Nyquist signaling}
\acro{GFDM}{generalized frequency division multiplexing}
\acro{GFDMA}{generalized frequency division multiple access}
\acro{GMC-CDM}{generalized	multicarrier code-division multiplexing}
\acro{GNSS}{global navigation satellite system}
\acro{GS}{guard symbols}
\acro{GSM}{Groupe Sp\'{e}cial Mobile}
\acro{GUI}{graphical user interface}
\acro{H2H}{human-to-human}
\acro{H2M}{human-to-machine}
\acro{HF}{high frequency}
\acro{HTC}{human type communication}
\acro{I}{in-phase}
\acro{i.i.d.}{independent and identically distributed}
\acro{IB}{in-band}
\acro{IBI}{inter-block interference}
\acro{IC}{interference cancellation}
\acro{ICI}{inter-carrier interference}
\acro{ICT}{information and communication technologies}
\acro{ICV}{information coefficient vector}
\acro{IDFT}{inverse discrete Fourier transform}
\acro{IDMA}{interleave division multiple access}
\acro{IEEE}{institute of electrical and electronics engineers}
\acro{IF}{intermediate frequency}
\acro{IFFT}{inverse fast Fourier transform}
\acro{IoT}{Internet of Things}
\acro{IOTA}{isotropic orthogonal transform algorithm}
\acro{IP}{internet protocole}
\acro{IP-core}{intellectual property core}
\acro{ISDB-T}{terrestrial integrated services digital broadcasting}
\acro{ISDN}{integrated services digital network}
\acro{ISI}{inter-symbol interference}
\acro{JCNS}{joint communication \& sensing}
\acro{ITU}{International Telecommunication Union}
\acro{IUI}{inter-user interference}
\acro{LAN}{local area netwrok}
\acro{LLR}{log-likelihood ratio}
\acro{LMMSE}{linear minimum mean square error}
\acro{LNA}{low noise amplifier}
\acro{LO}{local oscillator}
\acro{LOS}{line-of-sight}
\acro{LoS}{line of sight}
\acro{LP}{low-pass}
\acro{LPF}{low-pass filter}
\acro{LS}{least squares}
\acro{LTE}{long term evolution}
\acro{LTE-A}{LTE-Advanced}
\acro{LTIV}{linear time invariant}
\acro{LTV}{linear time variant}
\acro{LUT}{lookup table}
\acro{M2M}{machine-to-machine}
\acro{MA}{multiple access}
\acro{MAC}{multiple access control}
\acro{MAP}{maximum a posteriori}
\acro{MAE}{mean absolute error}
\acro{MC}{multicarrier}
\acro{MCA}{multicarrier access}
\acro{MCM}{multicarrier modulation}
\acro{MCS}{modulation coding scheme}
\acro{MF}{matched filter}
\acro{MF-SIC}{matched filter with successive interference cancellation}
\acro{MIMO}{multiple-input multiple-output}
\acro{MISO}{multiple-input single-output}
\acro{ML}{maximum likelihood}
\acro{MLD}{maximum likelihood detection}
\acro{MLE}{maximum likelihood estimator}
\acro{MMSE}{minimum mean squared error}
\acro{MRC}{maximum ratio combining}
\acro{MS}{mobile stations}
\acro{MSE}{mean squared error}
\acro{MSK}{Minimum-shift keying}
\acro{MSSS}[MSSS]	{mean-square signal separation}
\acro{MTC}{machine type communication}
\acro{MU}{multi user}
\acro{MVDR}{minimum variance distortionless response}
\acro{MVUE}{minimum variance unbiased estimator}
\acro{NEF}{noise enhancement factor}
\acro{NLOS}{non-line-of-sight}
\acro{NMSE}{normalized mean-squared error}
\acro{NOMA}{non-orthogonal multiple access}
\acro{NPR}{near-perfect reconstruction}
\acro{NRZ}{non-return-to-zero}
\acro{OU}{Ornstein-Uhlenbeck}
\acro{OFDM}{orthogonal frequency division multiplexing}
\acro{OFDMA}{orthogonal frequency division multiple access}
\acro{OMP}{orthogonal matching pursuit}
\acro{OOB}{out-of-band}
\acro{OQAM}{offset quadrature amplitude modulation}
\acro{OQPSK}{offset quadrature phase shift keying}
\acro{OTFS}{orthogonal time frequency space}
\acro{PA}{power amplifier}
\acro{PAM}{pulse amplitude modulation}
\acro{PAPR}{peak-to-average power ratio}
\acro{PB}{pass-band}
\acro{PC-CC}{parallel concatenated convolutional code}
\acro{PCP}{pseudo-circular pre/post-amble}
\acro{PD}{probability of detection}
\acro{pdf}{probability density function}
\acro{PDF}{probability distribution function}
\acro{PDP}{power delay profile}
\acro{PFA}{probability of false alarm}
\acro{PFD}{phase-frequency detector}
\acro{PHY}{physical layer}
\acro{PIC}{parallel interference cancellation}
\acro{PLC}{power line communication}
\acro{PLL}{phase-locked loop}
\acro{PMF}{probability mass function}
\acro{PN}{pseudo noise}
\acro{ppm}{parts per million}
\acro{PRB}{physical resource block}
\acro{PRB}{physical resource block}
\acro{PSD}{power spectral density}
\acro{PUSCH}{physical uplink shared channel}
\acro{Q}{quadrature-phase}
\acro{QAM}{quadrature amplitude modulation}
\acro{QoS}{quality of service}
\acro{QPSK}{quadrature phase shift keying}
\acro{R/W}{read-or-write}
\acro{RAM}{random-access memmory}
\acro{RAN}{radio access network}
\acro{RAT}{radio access technologies}
\acro{RC}{raised cosine}
\acro{RE}{resource element}
\acro{RF}{radio frequency}
\acro{RMS}{root mean square}
\acro{RMSE}{root mean square error}
\acro{RSSI}{received signal strength indicator}
\acro{RRC}{root raised cosine}
\acro{REF}{reference}
\acro{RW}{read-and-write}
\acro{RX}{receiver}
\acro{SC}{single-carrier}
\acro{SCA}{single-carrier access}
\acro{SC-FDE}{single-carrier with frequency domain equalization}
\acro{SC-FDM}{single-carrier frequency division multiplexing}
\acro{SC-FDMA}{single-carrier frequency division multiple access}
\acro{SD}{sphere decoding}
\acro{SDD}{space-division duplexing}
\acro{SDMA}{space division multiple access}
\acro{SDR}{software-defined radio}
\acro{SDW}{software-defined waveform}
\acro{SEFDM}{spectrally efficient frequency division multiplexing}
\acro{SE-FDM}{spectrally efficient frequency division multiplexing}
\acro{SER}{symbol error rate}
\acro{SIC}{successive interference cancellation}
\acro{SINR}{signal-to-interference-plus-noise ratio}
\acro{SIR}{signal-to-interference ratio}
\acro{SISO}{single-input, single-output}
\acro{SRS}{sounding reference signal}
\acro{SMS}{Short Message Service}
\acro{SNR}{signal-to-noise ratio}
\acro{STC}{space-time coding}
\acro{STFT}{short-time Fourier transform}
\acro{STO}{sample-time-offset}
\acro{SU}{single user}
\acro{SVD}{singular value decomposition}
\acro{TX}{transmitter}
\acro{TD}{time-domain}	
\acro{TDD}{time-division duplexing}
\acro{TTD}{true-time delay}
\acro{TDMA}{time-division multiple access}
\acro{TFL}{time-frequency localization}
\acro{TO}{time offset}
\acro{ToA}{time of arrival}
\acro{TDoA}{time difference of arrival}
\acro{TS-OQAM-GFDM}{time-shifted OQAM-GFDM}
\acro{UE}{user equipment}
\acro{UFMC}{universally filtered multicarrier}
\acro{UL}{uplink}
\acro{ULA}{uniform linear array}
\acro{US}{uncorrelated scattering}
\acro{USB}{universal serial bus}
\acro{USRP}{universal software radio peripheral}
\acro{UW}{unique word}
\acro{VLC}{visible light communications}
\acro{VR}{virtual reality}
\acro{VCO}{voltage-controlled oscillator}
\acro{WCP}{windowing and \acs{CP}}	
\acro{WHT}{Walsh-Hadamard transform}
\acro{WiMAX}{worldwide interoperability for microwave access}
\acro{WLAN}{wireless local area network}
\acro{W-OFDM}{windowed-\acs{OFDM}}	
\acro{WOLA}{windowing and overlapping}	
\acro{WSS}{wide-sense stationary}
\acro{ZCT}{Zadoff-Chu transform}
\acro{ZF}{zero-forcing}
\acro{ZMCSCG}{zero-mean circularly-symmetric complex Gaussian}
\acro{ZP}{zero-padding}
\acro{ZT}{zero-tail}
\acro{ZC}{Zadoff-Chu}
\acro{URLLC}{ultra-reliable low-latency communications}

\acro{HSI}{human system interface}
\acro{HMI}{human machine interface}
\acro{VR} {visual reality} 
\acro{AGV}{automated guided vehicles}
\acro{MEC}{multiaccess edge cloud}
\acro{TI} {tactile Internet}
\acro{IMT}{ international mobile telecommunications}
\acro{GN}{gateway node}
\acro{CN}{control node}
\acro{NC}{network controller}
\acro{SN}{sensor node}
\acro{AN}{actuator node}
\acro{HN}{haptic node}
\acro{TD}{tactile devices}
\acro{SE}{supporting engine}
\acro{AI}{artificial intelligence}
\acro{TSM}{tactile service manager}
\acro{TTI}{transmission time interval}
\acro{NR}{new radio}
\acro{SDN}{software defined networking}
\acro{NFV}{ network function virtualization}
\acro{CPS}{cyber-physical system}
\acro{TSN}{Time-Sensitive Networking}
\acro{FEC}{forward error correction}
\acro{STC}{space-time  coding}
\acro{HARQ}{hybrid automatic repeat request}
\acro{CoMP} {Coordinated multipoint}
\acro{HIS}{human system interface }
\acro{RU}{radio unit}
\acro{CU}{central unit}
\acro{AoD} {angle of departure}
\end{acronym}
    \title{Angle Estimation for Initial Access in True-Time-Delay Systems: Communication and Sensing Tradeoff}
	
	\author{
		\IEEEauthorblockN{
			Carl Collmann, Ahmad Nimr, Gerhard Fettweis
			}
			
		\IEEEauthorblockA{
		Vodafone Chair Mobile Communications Systems, Technische Universit\"{a}t Dresden, Germany\\ \small\texttt{\{carl.collmann, ahmad.nimr,  gerhard.fettweis\}@tu-dresden.de}\\
		}
		}
	\maketitle
	\IEEEpeerreviewmaketitle
	
\begin{abstract}

In \ac{JCNS} the data signal is used to obtain parameter estimates.
An example is the \ac{AoA} which can be used to solve the initial access problem of beam selection.
\ac{TTD} arrays are attractive for this purpose, as their frequency-dependent response encodes the angle information in the receive signal power spectrum.
This work investigates the tradeoff between angle estimation accuracy and achievable data rate for uplink signals in a \ac{TTD} array.
The \ac{FI} for the angle is used to find the \ac{BCRLB} under the assumption of a raised cosine distributed angle prior.
Maximization of the \ac{FI} under a data rate constraint is solved by waterfilling power per frequency bin dependent on sensing weights.
Compared to a frequency flat signal, the optimized power allocation increases the \ac{FI} $\approx\SI{4.9}{dB}$ at $5\%$ rate loss for prior width of $\pm \SI{10}{\degree}$.
Numerical simulations show that the derived \ac{BCRLB} can be attained by a \ac{ML} estimator above an \ac{SNR} threshold, confirming that the \ac{FI} underlying the waveform design is attainable.
\end{abstract}

\begin{IEEEkeywords}
true-time-delay, angle estimation, joint communication and sensing, waveform design 
\end{IEEEkeywords}
 
	\acresetall
\section{Introduction}\label{sec:introduction}


In 6G systems, \ac{JCNS} will play an important role by transforming cell infrastructures into perceptive networks \cite{fanliu_9737357}. 
Parameter estimates such as the \ac{AoA}, derived from sensing, are key enablers of smart, context-aware mobile communication systems \cite{Liu9652071}.
They can be used for beam selection, localization, object tracking, and for the synchronization and calibration of increasingly large antenna arrays in \ac{MIMO} systems \cite{rogalin6760595}.

\Ac{TTD} arrays apply a frequency-dependent phase shift, resulting in a frequency dependent beam.
Recently, they have regained interest for fast beam training \cite{Aditya9723402} in initial access procedures.
The property of frequency selectivity makes \ac{TTD} arrays attractive for sensing, as the \ac{AoA} information is contained in the frequency response of the combined receive signal.
Angle estimation methods for \ac{TTD} systems include \ac{OMP} and \ac{ML} based estimation \cite{2022JSPSy941015B}, \ac{LMMSE} \cite{jans9097132} or correlation- and model-based estimators \cite{Coll202403}.

In \ac{JCNS}, the waveform serves a dual purpose of communication and sensing.
Existing works on waveform design optimize the tradeoff between communication rate and sensing metrics such as the \ac{CRLB} \cite{Liu9652071}, and characterize the fundamental limits of this tradeoff for Gaussian channels \cite{Xiong10147248}.
In the case of \ac{TTD} arrays, the sensing information and communication performance depend both on the power allocation over frequency bins.
To the best of our knowledge this tradeoff has not been studied for \ac{TTD} arrays.
In an uplink sensing scenario, the \ac{BS} configures pilot and data allocation for the \ac{UE}, so that the waveform design choice is a scheduling decision at the \ac{BS}.


This work builds on the system model and the \ac{CRLB} for \ac{TTD} arrays in
\cite{collmann2026cramerraoboundanglearrival} to study this tradeoff.
The contributions are:
\begin{itemize}
  \item A Bayesian formulation of the sensing limit, which accounts for the
        angular uncertainty of the \ac{UE} through an angle prior.
  \item A convex formulation that maximizes the sensing information under a
        rate constraint. Its solution is a waterfilling in which the water
        level depends on the sensing weight of each frequency bin.
  \item An evaluation of the resulting rate--sensing tradeoff for a 5G NR
        uplink \ac{SRS} configuration
  \item A Monte-Carlo evaluation showing that the bound can be attained by an \ac{ML} estimator above a certain \ac{SNR} threshold and compared with a low-complexity peak estimator \cite{Coll202403}.
\end{itemize}

The paper is structured as follows:
Section \ref{sec:system_model} introduces the general system model.
Section~\ref{sec:crb} derives the Fisher information and the \ac{CRLB}.
Section~\ref{sec:wave_design} states the tradeoff problem and introduces the \ac{BCRLB}.
Section~\ref{sec:evaluation} evaluates the \ac{JCNS} tradeoff and shows attainability of the bound with a numerical simulation.
Section~\ref{sec:conclusions} concludes the paper.

    \begin{figure}[tb]
    	\centering
    	\includegraphics[width=\linewidth]{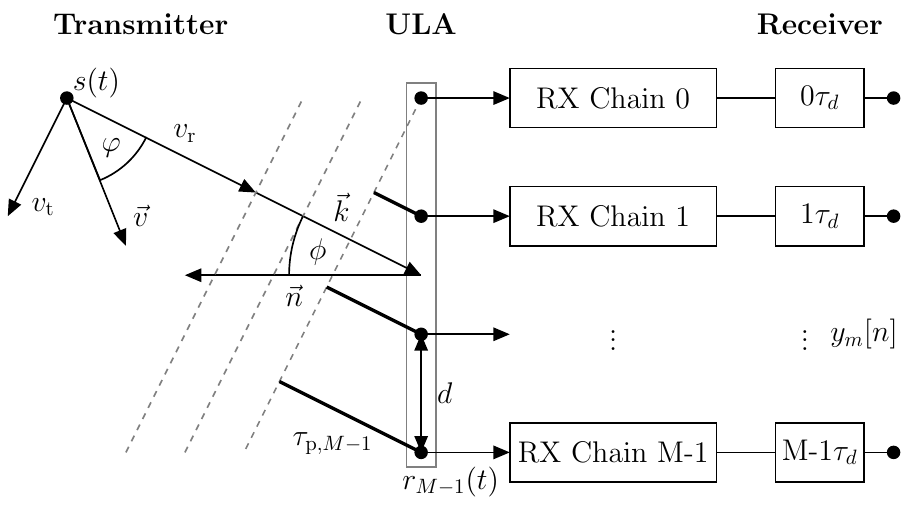}
        \caption{Schematic of an \ac{ULA} consisting of $M$ elements spaced by distance $d$. The vector $\vec{k}$ denotes the direction from the transmitter (\ac{TX}) to the receiver (\ac{RX}), while the vector $\vec{n}$ refers to the array normal.}
    	\label{fig:ula_array}
     \vspace{-5 mm}
\end{figure}

\section{System Model}\label{sec:system_model}
The considered system configuration is illustrated in \figref{fig:ula_array}.
A transmitter, representing a \ac{UE} communicating with a \ac{BS} in the uplink, omnidirectionally radiates a signal
\begin{align}
    s(t) = \mathfrak{Re} \{ x(t) e^{j2\pi f_\text{c} t}\}.
\end{align}
The \ac{BS} is equipped with a \ac{ULA} of $M$ elements spaced at $d=\lambda/2$ and with \ac{AoA} $\phi$ relative to the array normal.
%
The received signal at the $m$-th antenna element experiences a channel gain $\rho$ and a total delay $\tau_{\text{ch}}+\tau_{\text{p},m}$, where $\tau_{\text{ch}}$ denotes the common channel delay and $\tau_{\text{p},m} = m\frac{d}{c}\sin \phi$ represents the additional propagation delay associated with the $m$-th antenna element.
\begin{align}
    r_m(t) = \mathfrak{Re} \{ \rho x(t - \tau_\text{ch} - \tau_{\text{p},m}) e^{j2\pi f_\text{c}(t - \tau_\text{ch} - \tau_{\text{p},m})} \}.
\end{align}
After down-conversion the baseband signal delayed by true-time delay $\tau_{\text{d},m} = m\tau_d$ is expressed as
\begin{align}
    y_m(t) = \rho x(t - \tau_\text{ch} - \tau_{\text{p},m} - \tau_{\text{d},m}) e^{-j2\pi f_\text{c}(\tau_\text{ch} + \tau_{\text{p},m}) }.
\end{align}
By applying the Fourier transform, the $m$-th signal in the frequency domain is given by
\begin{align}
    Y_m(f) &= \rho X(f) e^{- j2\pi f(\tau_\text{ch} + \tau_{\text{p},m} + \tau_{\text{d},m})} e^{-j2\pi f_\text{c} (\tau_\text{ch} + \tau_{\text{p},m})}\nonumber\\
        &= \rho X(f) e^{-j2\pi (f_\text{c} + f)\tau_\text{ch}} e^{-jm\psi},
\end{align}
with $\psi(f,\phi) = 2\pi[(f_\text{c} + f)\frac{d}{c}\sin\phi + f \tau_{\text{d}}]$. The signals are combined to $Y(f) = \sum_{m=0}^{M-1} Y_m(f)$, and 
the system function $H=Y/X$ can be expressed as
\begin{align}
    H(f,\phi) = \rho' \sum_{m=0}^{M-1} e^{-jm\psi(f,\phi)},
\end{align}
where $\rho'=\rho e^{- 2\pi (f_\text{c} + f)\tau_\text{ch}}$.
Since for angle estimation a common phase shift applied to the signal is of no significance and the gain of the receive signal can be controlled with \ac{AGC}, for simplicity it is assumed that $\rho'$ is known and compensated , so that $\rho'=1$.

A noise free observation of the receive signal is 
\begin{align}
    Y(f,\phi) =  {X(f)} \underbrace{\sum_{m=0}^{M-1} e^{-jm2\pi[(f_\text{c} + f)\frac{d}{c}\sin\phi + f \tau_{\text{d}}]}}_{H(f,\phi)},
\end{align}
and noisy observation
\begin{align}
    Z(f) = X(f)H(f,\phi) + V(f),
    \label{eq:sys_model}
\end{align}
with Gaussian noise ${ V(f) \sim \mathcal{C}\mathcal{N}(0, \sigma_v^2)}$. 

\begin{figure}[tb]
    	\centering
    	\includegraphics[width=0.8\linewidth]{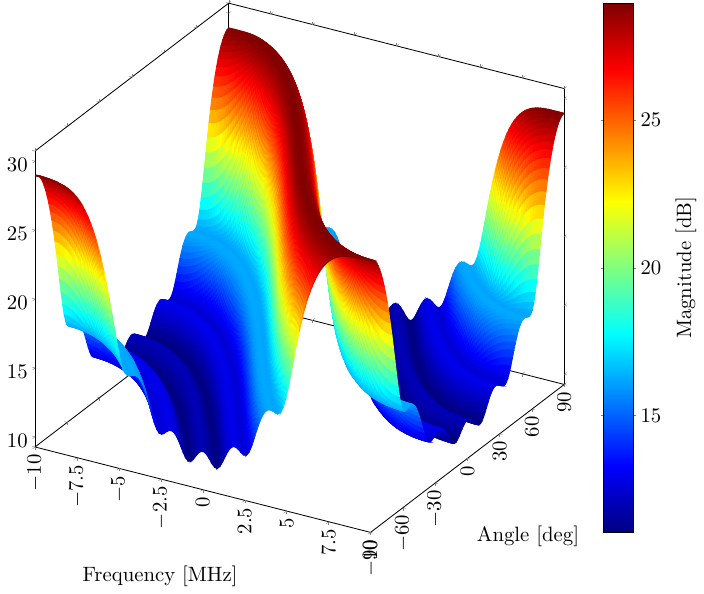}
        \caption{$10\log_\text{10}(\kappa(f,\phi))$ for $M=8,f_\text{c}= \SI{3.75}{GHz}, B=\SI{20}{MHz}, \tau_\text{d}=1/B$}
    	\label{fig:kappa_f_phi}
        \vspace{-5 mm}
\end{figure}
\section{Cramer-Rao Bound for AoA Estimates in TTD Systems}\label{sec:crb}

The objective is to estimate the angle $\phi$ from the received signal based on the previously established model.
The Cram\'er-Rao bound provides a lower bound on the variance of any unbiased estimator and therefore serves as a benchmark for evaluating angle estimation methods in \ac{TTD} systems.
While the derivation assumes no presence of hardware impairments, it has been shown that the angle estimation bounds can still be applicable for misspecified models \cite{Fortunati8103120}, given that their magnitude is small.

By considering $n$ discrete samples/frequency bins at frequencies $f=f_n$, the observation is
\begin{align}
    Z[n] = Y[n,\phi] + V[n] = Z_n,
\end{align}
with $Y[n,\phi]= X_n H(f_n, \phi)$, $X_n = X(f_n)$. 
The likelihood function for the noisy received signal observation vector $\boldsymbol{Z}$ is 
\begin{align}
    p(\boldsymbol{Z};\phi) = \prod_{n=1}^{N} \frac{1}{\pi \sigma_v^2} e^{-\frac{|Z_n - Y_n|^2}{\sigma_v^2}},
\end{align}
and the corresponding log-likelihood function
\begin{align}
    \mathcal{L}(\phi) = \ln p(\boldsymbol{Z};\phi) = -\ln{\pi \sigma_v^2} - \frac{1}{\sigma_v^2}\sum_{n=1}^{N} |Z_n-Y_n|^2.
\end{align}
The \ac{FI} is defined as
\begin{align}
    &\mathcal{I}(\phi) = - \mathbb{E}\left[ \frac{\partial^2 \mathcal{L}(\phi)}{\partial \phi^2} \right] = \frac{2}{\sigma_v^2}\sum_{n=1}^{N} \left|\frac{\partial Y_n}{\partial\phi} \right|^2\\
    &= \frac{2}{\sigma_v^2}\sum_{n=1}^{N} \left| X_n \left(-j2\pi (f_n + f_\text{c})\frac{d}{c}\cos{\phi}\right) \sum_{m=0}^{M-1} m e^{-jm\psi_n} \right|^2\nonumber\\
    &\le \frac{2}{\sigma_v^2}\sum_{n=1}^{N} \underbrace{\left| X_n \right|^2}_{\gamma_n} \cdot \left|2\pi (f_n + f_\text{c})\frac{d}{c}\cos{\phi}\right|^2 \cdot  \underbrace{ \left|\sum_{m=0}^{M-1} m e^{-jm\psi_n} \right|^2}_{\kappa_n}.\nonumber
    \label{eq:FI_full}
\end{align}
Here, $ \psi_n = 2\pi[(f_\text{c} + f_n)\frac{d}{c}\sin\phi + f_n \tau_{\text{d}}]$.
The sum over antenna elements $m$ can be rewritten using the property of a geometric series 
$ \sum_{m=0}^{M-1} e^{-jm\psi_n} = {(1-e^{-jM\psi_n})}/{(1-e^{-j\psi_n})}$ and taking its first derivative w.t.r $\psi_n$ results in 
\begin{align}
    \kappa_n(\phi) &= \left|\sum_{m=0}^{M-1} m e^{-jm\psi_n} \right|^2 = \kappa(f_n, \phi)\\
  \nonumber  &= \frac{\left| e^{-j\psi_n} - Me^{-jM\psi_n} + (M-1)e^{-j(M+1)\psi_n} \right|^2}{16 \sin^4(\psi_n/2)}
\end{align}
For illustration the variable $\kappa(f,\phi)$ is plotted in \figref{fig:kappa_f_phi}.
When selecting a carrier frequency $f_c=c/2d$ and \ac{TTD} $\tau_d=B^{-1}$, $\psi_n$ sweeps the full angular domain over the frequency band.
The term $\kappa_n$ can be understood as carrying information of angle-dependent frequency-scanning beam.
For frequency bins $f_n$ sampled uniformaly over  the interval $[-B/2, B/2]$ with spacing $\Delta f=B/N$, the \ac{FI} can be approximated by a Riemann integral, and therefore,  
\begin{align}
     \mathcal{I}(\phi) &\le \frac{2}{\sigma_v^2} \left( 2\pi \frac{d}{c} \cos\phi \right)^2\\
     &\cdot\frac{1}{\Delta f} \int_{-B/2}^{B/2} |X(f)|^2 (f+f_\text{c})^2 \kappa (f) df.\nonumber
\end{align}
Under the assumption that the spectrum of $X(f)$ is flat, i.e. $|X(f)|^2 = |X(0)|, ~ f\in [-B/2, B/2]$, and $f_\text{c} \gg B$ (narrowband), then 
\begin{align}
    \mathcal{I}(\phi)&\le \frac{2E_s \kappa_0(\phi)}{\sigma_v^2} \left( 2\pi f_\text{c} \frac{d}{c} \cos\phi \right)^2,
    \label{eq:FI_simpl}
\end{align}
with signal energy $E_s = \sum_{n} |X_n|^2 = N|X(0)|$ and 
\begin{equation}
\kappa_0(\phi) = \frac{1}{B}\int_{-B/2}^{B/2}\kappa (f, \phi)df    
\end{equation}
Then an approximation of the \ac{CRLB} for $\phi$ is given as
\begin{align}
    \text{Var}(\hat{\phi}) &\ge \frac{1}{\mathcal{I}(\phi)}\nonumber\\
    &\ge \frac{\sigma_v^2}{2E_s \kappa_0(\phi) \left( 2\pi f_\text{c} \frac{d}{c} \cos\phi \right)^2}.
    \label{eq:crlb_simpl}
\end{align}

    \section{Waveform Design}
\label{sec:wave_design}
\subsection{Problem Setup}

From \eqref{eq:FI_full} it can be seen that the \ac{FI} is linear in regards to the per-bin power $\gamma_n$
\begin{align}
    \mathcal{I}(\phi;\gamma) = \sum_{n=1}^{N} a_n(\phi) \gamma_n, \label{eq:fi_problem}
\end{align}
with
\begin{align}
    a_n(\phi) = \frac{2}{\sigma_v^2} \left(2\pi \frac{d}{c}\cos{\phi}\right)^2 (f_n + f_\text{c})^2 \kappa_n(\phi).
\end{align}
The sensing performance is dependent on the waveform, which opens the question of choosing a favourable waveform and allocation of frequency resources.
As the \ac{BS} is responsible for configuration of the \acs{UE} \ac{SRS} and \ac{PUSCH}, optimization of $\gamma_n = |X_n|^2$ is a scheduling decision the \ac{BS} should compute.

Since the signals at the receiver are combined after applying \acs{TTD}, the $n$-th bin is subjected to the gain
\begin{equation}
    |H_n(\phi)|^2 = \left|\sum_{m=0}^{M-1} e^{-jm\psi(f_n,\phi)}\right|^2 = \frac{\sin^2(M \psi_n/2)}{\sin^2(\psi_n/2)}.
\end{equation}

The rate is
\begin{align}
    \mathcal{R}(\gamma) = \sum_{n=1}^{N} \log_2 \left(1+ \frac{\gamma_n|H_n(\phi)|^2}{\sigma_v^2}\right).
    \label{eq:rate}
\end{align}
To improve the angle estimation, it is required to choose $\{\gamma_n\}$ that maximizes the Fisher information, while maintaining a minimum data rate under a fixed power constraint. Therefore, the design can be formulated as 
\begin{align}
    \underset{\gamma \geq 0}{\max}~ \mathcal{I}(\phi;\gamma)~\text{s.t.}~\sum_n \gamma_n \leq E_s,~\mathcal{R}(\gamma) \geq \mathcal{R}_\text{min} .
    \label{eq:max_fi}
\end{align}

\subsection{Waveform Design with Waterfilling}

The goal is to maximize \eqref{eq:max_fi} for an energy budget $\sum_n \gamma_n \leq E_s$, data rate requirement $\mathcal{R}(\gamma) \geq \mathcal{R}_\text{min}$ \cite{convex_opt}. 
To formulate the tradeoff between sensing and communication, a cost function is introduced with $\lambda$ scaling the cost for allocating power per bin and $\mu$ for scaling the impact of the communication in the distribution of power
\begin{align}
    \mathcal{J}_n(\gamma_n) = \underbrace{a_n(\phi) \gamma_n}_{FI,~sensing} + \underbrace{\mu \log_2 \left(1+ \frac{\gamma_n|H_n(\phi)|^2}{\sigma_v^2}\right)}_{rate,~communication} - \underbrace{\lambda \gamma_n}_{power}.
\end{align}
The sum over $\mathcal{J}_n$ yields the Lagrangian of the problem with constants $\mu \mathcal{R}_\text{min}$ and $\lambda E_s$.
Since the bins are decoupled, their optimal values can be optimized independently of others for a given set of $\lambda, \mu$.
As the function  $\mathcal{J}_n$ is concave in $\gamma_n$, its maximum can be found where its derivative becomes zero
\begin{align}
    \frac{\partial \mathcal{J}_n(\gamma_n)}{\partial \gamma_n} = a_n(\phi) - \lambda + \frac{\mu |H_n(\phi)|^2}{\sigma_v^2 \ln(2) (1+\frac{\gamma_n|H_n(\phi)|^2}{\sigma_v^2})} \overset{!}{=} 0.
    \label{eq:derivative_cost_func}
\end{align}
This results in the following solution
\begin{align}
    \gamma_n^* = \left[ \frac{\mu}{\ln(2) (\lambda - a_n(\phi))} - \frac{\sigma_v^2}{|H_n(\phi)|^2}\right]^+.
    \label{eq:tilted_waterfill}
\end{align}
It is required that $\lambda > \underset{n}{\text{max}}~a_n$ to ensure that the power cost per bin exceeds the sensing value.

\subsection{Bayesian Approximation of Rate/CRLB Tradeoff}

The solution to \eqref{eq:tilted_waterfill} holds exactly only for a known angle.
To evaluate the expression for different angles, a Bayesian approach is considered.
Using Van Trees inequality \cite{bayes_bounds_trees2} yields
\begin{align}
    \mathbb{E}[(\hat\phi -\phi)^2] \geq [ \bar{\mathcal{I}}(\gamma) + \mathcal{I}_p]^{-1},
    \label{eq:bayesian_crlb}
\end{align}
with \ac{FI} average over angle $\bar{\mathcal{I}}(\gamma) = \sum_{n=1}^{N} \bar{a}_n \gamma_n$ and $\bar{a}_n = \mathbb{E}_\phi [{a}_n(\phi)]$ expectation in regards to angle, while $\mathcal{I}_p$ represents the priors \ac{FI}.
To comply with regularity conditions, a normalized raised cosine window of width $w$ centered around $\phi=\SI{0}{\degree}$ is selected as smooth prior.
Then the \ac{BCRLB} from \eqref{eq:bayesian_crlb} can be written as
\begin{align}
    &\mathbb{E}[(\hat\phi -\phi)^2] \geq \label{eq:bayesian_crlb2}\\ &\frac{\sigma_v^2}{ \frac{4\pi^2\sigma_v^2}{w^2} + 8 \left(\pi \frac{d}{c}\right)^2 \sum_{n=1}^N \gamma_n (f_n+f_\text{c})^2 \mathbb{E}_\phi[\text{cos}^2(\phi) \kappa_n(\phi)]}.\nonumber
\end{align}
The optimization result \eqref{eq:tilted_waterfill} is approximated for the Bayesian approach by simplifying the rate derivative in \eqref{eq:derivative_cost_func} so that
\begin{align}
    \gamma_n^* = \left[ \frac{\mu}{\lambda - \mathbb{E}_\phi [a_n(\phi)]} - \frac{\sigma_v^2}{ \mathbb{E}_\phi [|H_n(\phi)|^2]}\right]^+.
    \label{eq:tilted_waterfill_bayes}
\end{align}

The tradeoff between sensing and communication, expressed through the parameters $\lambda,\mu$, is found numerically.
Each point of the tradeoff curve \figref{fig:CRLB_rate} corresponds to one value of $\mu$, and the sweep runs from high to low values of $\mu$.
For a fixed $\mu$:
\begin{itemize}
    \item For a trial $\lambda>\max_n \bar a_n$, the power of each bin is found
          independently: $\gamma_n=0$ if the left side of
          \eqref{eq:bayes_stationarity} evaluated at $\gamma_n=0$ does not
          exceed $\lambda$, otherwise $\gamma_n$ is the root of
          \begin{align}
              \bar a_n + \frac{\mu}{\ln 2}\,
              \mathbb E_\phi\!\left[\frac{|H_n(\phi)|^2/\sigma_v^2}
              {1+\gamma_n|H_n(\phi)|^2/\sigma_v^2}\right]=\lambda,
              \label{eq:bayes_stationarity}
          \end{align}
          which is found by bisection, as the left side decreases with $\gamma_n$.
          For a known angle, this reduces to \eqref{eq:tilted_waterfill}.
    \item Since $\sum_n\gamma_n$ decreases with $\lambda$, bisection over
          $\lambda$ finds the value for which $\sum_n\gamma_n=E_s$. This yields
          $\gamma_n^\ast$.
    \item The average expected rate is
          $\mathcal R=\mathbb E_\phi\big[\mathcal R(\gamma^\ast)\big]$ with
          $\mathcal R(\gamma^\ast)$ from \eqref{eq:rate}.
    \item The average \ac{FI} is
          $\bar{\mathcal I}=\sum_{n=1}^{N}\bar a_n\gamma_n^\ast$, with
          $\bar a_n=\mathbb E_\phi[a_n(\phi)]$.
\end{itemize}
Each $\mu$ yields the solution of the rate-constrained problem for $\mathcal R_\text{min}=\mathcal R$.
Since the problem \eqref{eq:max_fi} is convex, this solution is globally optimal, so the sweep over $\mu$ traces the same curve as a sweep over $\mathcal R_\text{min}$. 

For a known angle or angle prior, \eqref{eq:tilted_waterfill} and \eqref{eq:tilted_waterfill_bayes} represent waterfilling at the frequency bin level, where the floor is set by the inverse channel gain $\sigma_v^2/|H_n(\phi)|^2$.
For the angle prior, \eqref{eq:bayes_stationarity} has the same structure with $\bar a_n$ and an effective floor determined by the expectation.
Bins that carry high angle information receive more power than in classic waterfilling, while bins at nulls of the array ($|H_n|^2\approx 0$) receive none.
For the case of communication only $\mu \rightarrow \infty$, $\lambda$ scales with $\mu$ so that the influence of $a_n(\phi)$ becomes negligible and \eqref{eq:tilted_waterfill} reduces to classical waterfilling \cite{munz1188165}.
In the case of sensing only $\mu \rightarrow 0$, all power goes to $\underset{n}{\text{max}}~a_n(\phi)$.

\section{Evaluation}\label{sec:evaluation}

\begin{figure}[tb]
    \centering
    \includegraphics[width=\linewidth]{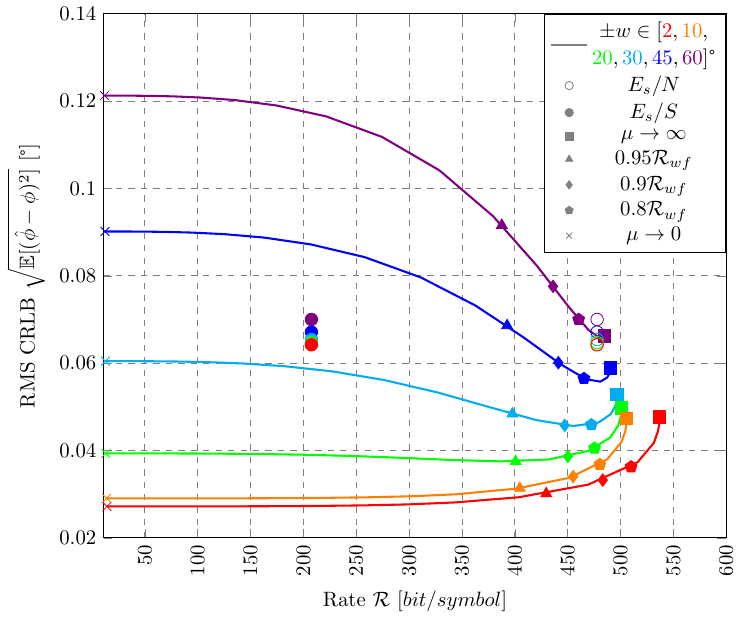}
    \caption{Tradeoff between expected data rate plotted over average \ac{CRLB} for different prior widths for $\text{SNR}=\SI{0}{dB}$.}
    \label{fig:CRLB_rate}
\end{figure}
 
\begin{table}[tb]
    \centering
    \begin{center}
        \caption{System parameters.}
        \vspace{-2mm}
            \label{table:system_parameters}
        \resizebox{0.8\linewidth}{!}{
        \begin{tabular}{| l | l | } 
         \hline
         Parameter & Value \\
         \hline
         Carrier freq. & $f_c=\SI{3.75}{GHz}$ \\ 
         SCS & $\Delta f_\text{SC} =\SI{30}{kHz}$\\
         FFT size & $N_\text{FFT}=1024$\\
         Sampl. rate & $f_s=\SI{30}{MHz}$\\
         Carrier grid & $51$ PRB's\\
         Cyclic prefix & $72$ Samples\\
         SRS allocation & 48 PRB start PRB 1\\
         SRS comb & $K_\text{TC}=2$\\
         SRS RE count/spacing & $288,~\SI{60}{kHz}$\\
         SRS used bandw. & $B=N\Delta f = \SI{17.28}{MHz}$\\
         SRS flat seq. & ZC-based, $N_\text{ZC}=283,u=7$\\
         SRS OFDM symb. & $1$\\
         ULA elements & $M=[8,16,32]$\\
         ULA spacing & $d=\SI{4}{cm}$\\
         True time delay & $\tau_d = \SI{57.9}{ns}$\\
         Noise variance & $\sigma_v^2 \sim \text{SNR}$\\
         SNR values & $E_s/(N \sigma_v^2) \in [-30, ..., 20]\SI{}{dB}$\\
         Angle prior $\phi_0 \pm w$ & $0 \pm [2,10,20,30,45,60]\SI{}{\degree}$\\
         Prior shape & raised cosine window\\
         \hline
        \end{tabular}
    }
    \end{center}
    \vspace{-5mm}
\end{table}
To illustrate the optimal tradeoff between communication and sensing performance, a sweep over $\mathcal{R}_\text{min}$ is performed and plotted against the \ac{RMS} of the \ac{CRLB} for different values of the prior width in \figref{fig:CRLB_rate}.
Note that with \eqref{eq:tilted_waterfill_bayes} the average \ac{FI} $\bar{\mathcal{I}}=\mathbb{E}_\phi[\mathcal{I}]$ is optimized, however the plot shows the average \ac{CRLB} $\sqrt{\mathbb{E}_\phi[1/\mathcal{I}]}$.
They are connected via Jensens inequality $\mathbb{E}[1/\mathcal{I}]\geq 1/\mathbb{E}[\mathcal{I}]$.
The rate counts the \acp{RE} as data, as the \ac{SRS} carries no data, its rate is a surrogate.
For a target rate $\mathcal{R}$, the curve gives an approximation of the largest average \ac{FI} that any power allocation $\gamma_n$ can achieve at that rate.
System parameters are given in \tabref{table:system_parameters}.
The following waveforms are considered:
\begin{itemize}
    \item \ac{SRS} based frequency flat \ac{ZC} sequence with $\gamma_n = E_s/S$ on comb $S$ with \ac{PRB} allocation according to \cite{3gpp_nr_ts_38_211}.
    \item Frequency flat \ac{OFDM} with $\gamma_n = E_s/N$ for all \ac{RE}.
    \item \ac{OFDM} waterfilled with $\gamma_n \sim$ \eqref{eq:tilted_waterfill} for $\mu \rightarrow \infty$ (maximum achievable data rate with waterfilling $\mathcal{R}_\text{wf}$).
    \item \ac{OFDM} modified waterfilling with $\gamma_n \sim$ \eqref{eq:tilted_waterfill} and $\mathcal{R}_\text{min} = \eta \mathcal{R}_\text{wf} \in [0.8,~0.9,~0.95]$
    \item Single tone at \ac{RE} where $\bar{a}_n$ is maximal with $\gamma_n \sim$ \eqref{eq:tilted_waterfill} for $\mu \rightarrow 0$ as limit point.
\end{itemize}

\subsection{Evaluation of Power Allocation with Waterfilling}

The following observations can be made in \figref{fig:CRLB_rate}:
\begin{itemize}
    \item Smaller prior width results in higher accuracy in angle estimates, independent of waveform.
    \item Neither frequency flat power allocation or frequency comb are optimal in regards of sensing accuracy or communication performance.
    \item The (modified) waterfilling approach yields power allocations $\gamma_n$ that are optimal in regards of maximizing sensing accuracy for a given data rate. Adjusting the minimum data rate with $\eta$ as a constraint when solving \eqref{eq:tilted_waterfill} can yield improved sensing accuracy.
    \item For small priors $\leq \SI{20}{\degree}$, higher data rate comes at a cost of less sensing accuracy in waveform design.
    \item For larger priors $\geq \SI{30}{\degree}$, this relation does not hold anymore and instead less data rate coincides with worse sensing performance. This can be explained by $a_n(\phi)$ having a similar shape to $|H(\phi)|$, so that for prior widths where spatial selectivity of both functions is significant, optimal data rate coincides with optimal sensing accuracy. 
\end{itemize}

\begin{figure}[tb]
    \centering
    \includegraphics[width=\linewidth]{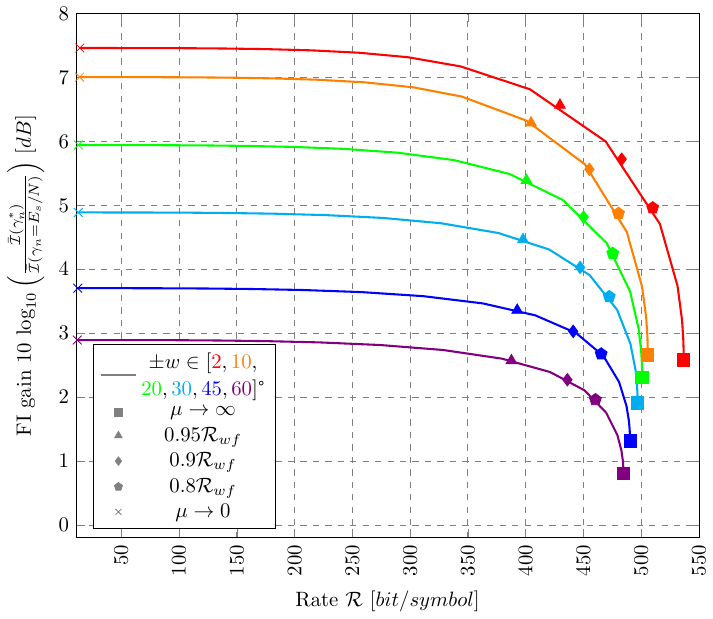}
    \caption{Improvement in \ac{FI} for waterfilling compared to even power allocation for different prior widths for $\text{SNR}=\SI{0}{dB}$.}
    \label{fig:fi_gain}
\end{figure}

The gain in \ac{FI} from using the proposed waterfilling approach compared to even power allocation between bins at different data rates is illustrated in \figref{fig:fi_gain} for different prior widths.
It can be seen that waterfilling always increases the \ac{FI}, regardless of the minimum rate used as a constraint for solving \eqref{eq:tilted_waterfill}.
This is explained by the similarity of $a_n(\phi)$ and $|H(\phi)|$, where the spacial selectivity favourable to communication also favours sensing accuracy.
A decent tradeoff between data rate and \ac{FI} gain is to take $\eta=0.8$ of the maximal rate $\mathcal{R}_{wf}$ which is close to the flat region for all curves.
In the flat region, further reduction in minimal data rate for the waveform design do not yield further improvements in \ac{FI}.
For small prior widths of $\pm \SI{2}{\degree}$, adjusting the minimal rate from $\mathcal{R}_{wf}$ to $0.8\mathcal{R}_{wf}$ provides an improvement of $\SI{4}{dB}$.
For large prior widths of $\pm \SI{60}{\degree}$, adjusting $\mathcal{R}_{wf}$ to $0.8\mathcal{R}_{wf}$ provides a smaller improvement of $\SI{1.8}{dB}$.
The general rule of higher prior width resulting in less \ac{FI} gain is explained that larger priors automatically result in a larger margin of error when optimizing $\gamma_n$ for a certain angle.

This motivates a two-stage sensing scheme where a frequency flat waveform first provides coarse estimate that serves as a narrow prior which is utilized by a waveform shaped by waterfilling in the second stage to maximize angle estimation accuracy by realizing the demonstrated \ac{FI} gain.

\begin{figure}[tb]
    \centering
    \includegraphics[width=\linewidth]{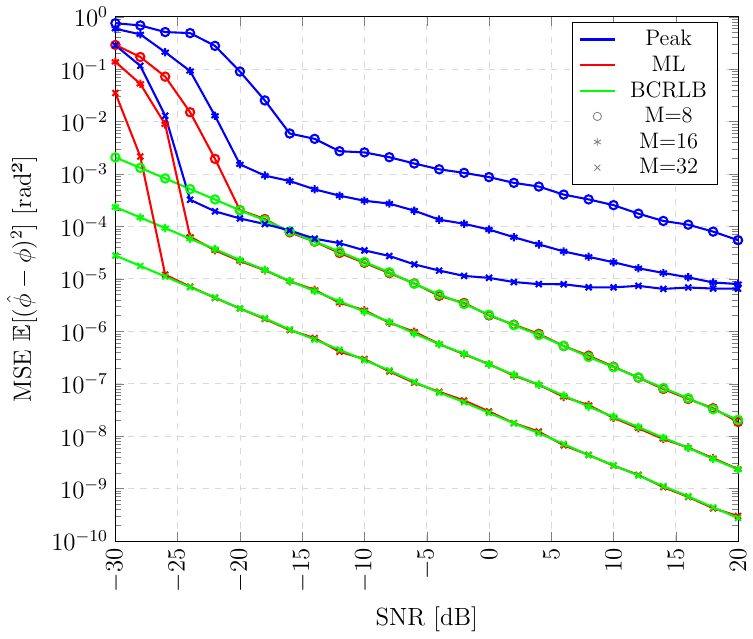}
    \caption{\ac{MSE} for \ac{ML} and Peak estimators at different \ac{SNR} compared to the numerically computed \ac{BCRLB}.}
    \label{fig:mse_snr}
\end{figure}

\subsection{Evaluating the Limit on Sensing Performance}

To ensure that the sensing limit imposed by the \ac{FI} \eqref{eq:fi_problem} which underlies the waveform design is attainable, the \ac{MSE} of two estimators is compared to the \ac{BCRLB} \eqref{eq:bayesian_crlb2}.
The angle prior is assumed to be a raised cosine window of width $w=2\pi/3$.
The observation signal vector $Z_n$ is generated according to the system model \eqref{eq:sys_model}, with a \ac{ZC} waveform and parameters \tabref{table:system_parameters}.
\ac{SNR} is defined as average \ac{RE} \ac{SNR} before array gain.

\emph{ML estimator:} For a known signal $X_n$, the \ac{ML} estimate of the angle is
\begin{align}
    \hat{\phi}_\text{ML} &= \underset{\phi}{\text{argmax}}~ \mathcal{L}(\phi) \\
    & = \underset{\phi}{\text{argmin}} \left[ \sum_{n=1}^{N} |Z_n-X_nH_n(\phi)|^2 \right].\nonumber
\end{align}
\emph{Peak estimator:} The observation $Z_n$ is smoothed by circular convolution with a rectangular window of $b=10$ frequency bins to enhance reliability of the estimator at low \ac{SNR} at the cost of angular resolution.
The frequency component with the maximum power $f^*$ is used in
\begin{align}
    \hat\phi_\text{PEAK}(f^*) = \text{arcsin} \left( - \frac{f^* \tau_d}{f^* + f_c} \frac{c}{d} \right),
    \label{eq:arcsin_map}
\end{align}
to obtain the angle estimate \cite{Coll202403}.

\figref{fig:mse_snr} shows the \ac{MSE} in relation to \ac{SNR} for the two considered estimators obtained from a Monte-Carlo simulation with $1000$ trials, for antenna arrays of size $M\in[8,16,32]$.
To judge the estimator accuracy they are compared to the numerically computed \ac{BCRLB} \eqref{eq:bayesian_crlb2} which is obtained by averaging over the angle prior.
It can be seen that the \ac{ML} estimator shows good agreement with the bound up until \ac{SNR} $\in-[20,24,26]~\SI{}{dB}$ for the selected antenna configuration, from which point on the bound is no longer attained.
The Peak estimator does not attain the \ac{BCRLB} and seems to be limited in angular resolution; the \ac{MSE} $\approx 10^{-5}~\SI{}{rad^2}$ doesn't improve from \ac{SNR} $\geq \SI{0}{dB}$ at $M=32$.
The reason for this effect lies in the limitation of this estimator by the number and spacing of frequency bins and also the smoothing applied to the observations.
Still, at $\SI{0}{dB}$ the Peak estimator obtains \ac{MSE} $\approx[10^{-3},10^{-4},10^{-5}]~\SI{}{rad^2}$ for the selected antenna configuration, which is sufficient for selecting a suitable beamformer at the \ac{BS} and solving the initial access problem.
The tradeoff in angle estimation accuracy for the Peak estimator compared to \ac{ML} can be justified with reduced computational complexity of applying \eqref{eq:arcsin_map} over a grid search for \ac{ML}, while an evaluation of this tradeoff is out of the scope of this work.
The main motivation of demonstrating that the \ac{FI} underlying the waveform design and consecutive sensing/communication tradeoff \eqref{eq:tilted_waterfill_bayes} is attainable, is verified by a \ac{ML} estimator.

\section{Conclusion}\label{sec:conclusions}


 
This work investigated the tradeoff between communication (data rate) and sensing (angle estimation accuracy) for uplink signals received by a \ac{TTD} array.
Because the \ac{FI} is linear in signal power, maximizing it under a rate constraint is a convex problem solved by waterfilling the signal power across frequency bins.
The \ac{FI} gain from waterfilling is inversely proportional to the prior angular width e.g., $\SI{7.4}{dB}$ for $w \in \pm \SI{2}{\degree}$.

Numerical simulations show that the derived \ac{BCRLB} can be attained by a \ac{ML} estimator up to \ac{SNR} thresholds of $-[20,24,26]~\SI{}{dB}$ for $M\in[8,16,32]$, validating the waterfilling based waveform design.
The Peak estimator is limited in angular resolution and does not attain the \ac{BCRLB}, but its \ac{MSE} is sufficient for beam selection in initial access.

Future work includes numerical complexity analysis and accuracy tradeoff for both estimators, detailed analysis of the proposed two stage design to refine the power allocation after a coarse initial estimate, and extensions for unknown data symbols and hardware impairments.

    \newpage
    \bibliographystyle{IEEEtran}
	\bibliography{references}

@STRING{TCOM="{IEEE Trans. Commun.}"}

@STRING{SPM="{IEEE Signal Processing Mag.}"}

@STRING{AC="{IEEE Trans. Autom. Control.}"}

@misc{collmann2026cramerraoboundanglearrival,
      title={Cramer-Rao Bound for Angle of Arrival Estimates in True-Time-Delay Systems}, 
      author={Carl Collmann and Ahmad Nimr and Gerhard Fettweis},
      year={2026},
      eprint={2601.03735},
      archivePrefix={arXiv},
      primaryClass={eess.SP},
      url={https://arxiv.org/abs/2601.03735}, 
}

@techreport{3gpp_nr_ts_38_211,
	author = {ETSI 3GPP},	
    institution = {},
	month = {09},
	note = {V19.5.0},
	number = {38.211},
	title = {{TS 38.211 5G; NR; Physical channels and modulation}},
	type = {Technical Specification (TS)},
	year = {2026}
}

@book{bayes_bounds_trees2,
    author = {Van Trees, H. and Bell, K.},
    title = {Bayesian Bounds for Parameter Estimation and Nonlinear Filtering/Tracking},
    publisher = {Wiley-IEEE Press},
    year = {2007}
}

@book{convex_opt,
    author = {Boyd, S. and Vandenberghe, L.},
    title = {Convex Optimization, Example 5.2},
    publisher = {Cambridge University Press},
    year = {2004}
}

@INPROCEEDINGS{munz1188165,
  author={Munz, G. and Pfletschinger, S. and Speidel, J.},
  booktitle={Global Telecommunications Conference, 2002. GLOBECOM '02. IEEE}, 
  title={An efficient waterfilling algorithm for multiple access OFDM}, 
  year={2002},
  volume={1},
  number={},
  pages={681-685 vol.1},
  doi={10.1109/GLOCOM.2002.1188165}}

@ARTICLE{Xiong10147248,
  author={Xiong, Yifeng and Liu, Fan and Cui, Yuanhao and Yuan, Weijie and Han, Tony Xiao and Caire, Giuseppe},
  journal={IEEE Transactions on Information Theory}, 
  title={On the Fundamental Tradeoff of Integrated Sensing and Communications Under Gaussian Channels}, 
  year={2023},
  volume={69},
  number={9},
  pages={5723-5751},
  doi={10.1109/TIT.2023.3284449}}

@ARTICLE{Liu9652071,
  author={Liu, Fan and Liu, Ya-Feng and Li, Ang and Masouros, Christos and Eldar, Yonina C.},
  journal={IEEE Transactions on Signal Processing}, 
  title={Cram\'{e}r-Rao Bound Optimization for Joint Radar-Communication Beamforming}, 
  year={2022},
  volume={70},
  number={},
  pages={240-253},
  doi={10.1109/TSP.2021.3135692}}

@ARTICLE{Fortunati8103120,
  author={Fortunati, Stefano and Gini, Fulvio and Greco, Maria S. and Richmond, Christ D.},
  journal={IEEE Signal Processing Magazine}, 
  title={Performance Bounds for Parameter Estimation under Misspecified Models: Fundamental Findings and Applications}, 
  year={2017},
  volume={34},
  number={6},
  pages={142-157},
  doi={10.1109/MSP.2017.2738017}}

@INPROCEEDINGS{Aditya9723402,
  author={Wadaskar, Aditya and Boljanovic, Veljko and Yan, Han and Cabric, Danijela},
  booktitle={2021 55th Asilomar Conference on Signals, Systems, and Computers}, 
  title={3D Rainbow Beam Design for Fast Beam Training with True-Time-Delay Arrays in Wideband Millimeter-Wave Systems}, 
  year={2021},
  volume={},
  number={},
  pages={85-92},
  doi={10.1109/IEEECONF53345.2021.9723402}}

@ARTICLE{rogalin6760595,
  author={Rogalin, R. and Bursalioglu, O. Y. and Papadopoulos, H. and Caire, G. and Molisch, A. F. and Michaloliakos, A. and Balan, V. and Psounis, K.},
  journal={IEEE Transactions on Wireless Communications}, 
  title={Scalable Synchronization and Reciprocity Calibration for Distributed Multiuser MIMO}, 
  year={2014},
  volume={13},
  number={4},
  pages={1815-1831},
  doi={10.1109/TWC.2014.030314.130474}}

@ARTICLE{fanliu_9737357,
  author={Liu, Fan and Cui, Yuanhao and Masouros, Christos and Xu, Jie and Han, Tony Xiao and Eldar, Yonina C. and Buzzi, Stefano},
  journal={IEEE Journal on Selected Areas in Communications}, 
  title={Integrated Sensing and Communications: Toward Dual-Functional Wireless Networks for 6G and Beyond}, 
  year={2022},
  volume={40},
  number={6},
  pages={1728-1767},
  doi={10.1109/JSAC.2022.3156632}}

@ARTICLE{2022JSPSy941015B,
       author = {{Boljanovic}, Veljko and {Cabric}, Danijela},
        title = "{Millimeter-Wave Wideband Channel Estimation Using Analog True-Time-Delay Array Under Hardware Impairments}",
      journal = {Journal of Signal Processing Systems},
         year = 2022,
        month = oct,
       volume = {94},
       number = {10},
        pages = {1015-1030},
          doi = {10.1007/s11265-022-01771-6},
       adsurl = {https://ui.adsabs.harvard.edu/abs/2022JSPSy..94.1015B}
}

@INPROCEEDINGS{Coll202403,
	author={C. Collmann and A.B. Martinez and A. Nimr and G. Fettweis},
	title={Angle-Of-Departure Estimation Algorithms for True-Time-Delay Antenna Arrays with Software-Defined Radio},
	booktitle={JC\&S 2024},
	month={Mar},
	year={2024},
	address={Leuven, Belgium},
}

@INPROCEEDINGS{jans9097132,
  author={Jans, Christoph and Song, Xiaohang and Rave, Wolfgang and Fettweis, Gerhard},
  booktitle={WSA 2020; 24th International ITG Workshop on Smart Antennas}, 
  title={Fast Beam Alignment through Simultaneous Beam Steering and Power Spectrum Estimation Using a Frequency Scanning Array}, 
  year={2020},
  volume={},
  number={},
  pages={1-6},
  doi={}}

@ARTICLE{MCM, 
	author={J. A. C. Bingham}, 
	journal={IEEE Communications Magazine}, 
	title={Multicarrier modulation for data transmission: an idea whose time has come}, 
	year={1990}, 
	volume={28}, 
	number={5}, 
	pages={5-14}, 
	doi={10.1109/35.54342}, 
	ISSN={0163-6804}, 
	month={May},}

@ARTICLE{FMT, 
	author={G. Cherubini and others}, 
	journal={IEEE Journal on Selected Areas in Communications}, 
	title={Filtered multitone modulation for very high-speed digital subscriber lines}, 
	year={2002}, 
	volume={20}, 
	number={5}, 
	pages={1016-1028}, 
	doi={10.1109/JSAC.2002.1007382}, 
	ISSN={0733-8716}, 
	month={June},}

@ARTICLE{OQAM, 
	author={I. Gaspar and others}, 
	journal={IEEE Communications Letters}, 
	title={{Frequency-Shift Offset-QAM for GFDM}}, 
	year={2015}, 
	volume={19}, 
	number={8}, 
	pages={1454-1457}, 
	doi={10.1109/LCOMM.2015.2445334}, 
	ISSN={1089-7798}, 
	month={Aug},}

@ARTICLE{OFDM, 
	author={T. Hwang and others}, 
	journal={IEEE Transactions on Vehicular Technology}, 
	title={{OFDM} and Its Wireless Applications: A Survey}, 
	year={2009}, 
	volume={58}, 
	number={4}, 
	pages={1673-1694}, 
	doi={10.1109/TVT.2008.2004555}, 
	ISSN={0018-9545}, 
	month={May},}

@article{FBMC,
	title={{{OFDM} versus filter bank multicarrier}},
	author={Farhang-Boroujeny, Behrouz},
	journal={SPM},
	volume={28},
	number={3},
	pages={92--112},
	year={2011},
	publisher={IEEE}
}

@ARTICLE{GFDM,
author={Michailow, Nicola and others},
	journal=TCOM,
	title={{Generalized Frequency Division Multiplexing for 5th Generation Cellular Networks}},
	year={2014},
	month=sep,
	volume={62},
	number={9},
	pages={3045-3061}, }

@INPROCEEDINGS{FDE, 
	author={A. Farhang and others}, 
	booktitle={2015 IEEE International Conference on Communications (ICC)}, 
	title={{Low complexity {GFDM} receiver design: A new approach}}, 
	year={2015}, 
	pages={4775-4780}, 
	doi={10.1109/ICC.2015.7249078}, 
	ISSN={1550-3607}, 
	month={June},
	address = {London, UK}}

@article{SVD,
	title={The singular value decomposition: Its computation and some applications},
	author={Klema, Virginia and Laub, Alan},
	journal=AC,
	volume={25},
	number={2},
	pages={164--176},
	year={1980},
	publisher={IEEE}
}

@inproceedings{Receiver,
	title={Generalized frequency division multiplexing: A flexible multi-carrier modulation scheme for 5th generation cellular networks},
	author={Michailow, Nicola and others},
	booktitle={Proceedings of the German microwave conference (GeMiC 12)},
	pages={1--4},
	year={2012},
	address = {Ilmenau, Germany}
}

\end{document}